\documentclass[12pt]{iopart}
\usepackage{iopams}
\usepackage{graphicx}
\begin{document}

\title{Light Nuclei Solving the AUGER puzzles.\\ The Cen-A 	
imprint.}
\author{D. Fargion}
\emph{}
\address{Physics Department and INFN, Rome University Sapienza, Pl.e A. Moro 2, 00185, Rome, Italy.}

\ead{daniele.fargion@roma1.infn.it}

\begin{abstract}
Ultra High Energy Cosmic Rays (UHECR) map at $60$ EeV have been
found recently by AUGER group spreading anisotropy signatures in the sky. The
result have been interpreted as a manifestation of AGN sources
ejecting protons at GZK edges, around or below $80$ Mpc distances, mostly from Super-galactic Plane. The result is surprising due to the lack of correlation
with much nearer Virgo cluster. Moreover, early GZK cut off in the spectra may be
better reconcile with light nuclei (than with protons). In addition a large group   (nearly a dozen) of events cluster suspiciously along  Cen-A.  Finally, proton  UHECR composition nature is in sharp disagreement with earlier AUGER claim of a  heavy nuclei dominance at $40$ EeV, within $13$ extreme events ($\ln A=2.6\pm0.6$).   Therefore, we interpret here the signals as \emph{mostly UHECR light nuclei (He, Be, B, C, O) ejected from nearest Cen-A, UHECR smeared by galactic magnetic   fields, whose random vertical bending is overlapping with super-galactic arm}. The (possible) AUGER  misunderstanding took place because of a rare coincidence between the Super  Galactic Plane (arm) and the smeared (randomized) signals from Cen-A, bent orthogonally to the Galactic fields. Our  derivation verify  the consistence of the random smearing angles  for He, Be, B, C, O range respectively $\gtrsim2.7^\circ-11^\circ$  in reasonable agreement with the AUGER main group event around Cen-A.  Only  few other rare events are spread elsewhere.  The most collimated  from Cen-A, the lightest ($\ln A_{He}\leq2$). The most spread the heavier ($\ln A\geq2$). Consequently Cen-A is probably one of the best candidate UHE neutrino  at tens-hundreds PeVs. This solution maybe tested soon by future (and maybe already recorded) clustering around the Cen-A barycenter, events smeared by vertical galactic magnetic forces on lightest nuclei.
\end{abstract}

\section{Introduction: Puzzled by AUGER puzzle}
\begin{figure}[t]
\begin{center}
\includegraphics[width=9cm, height=6cm]{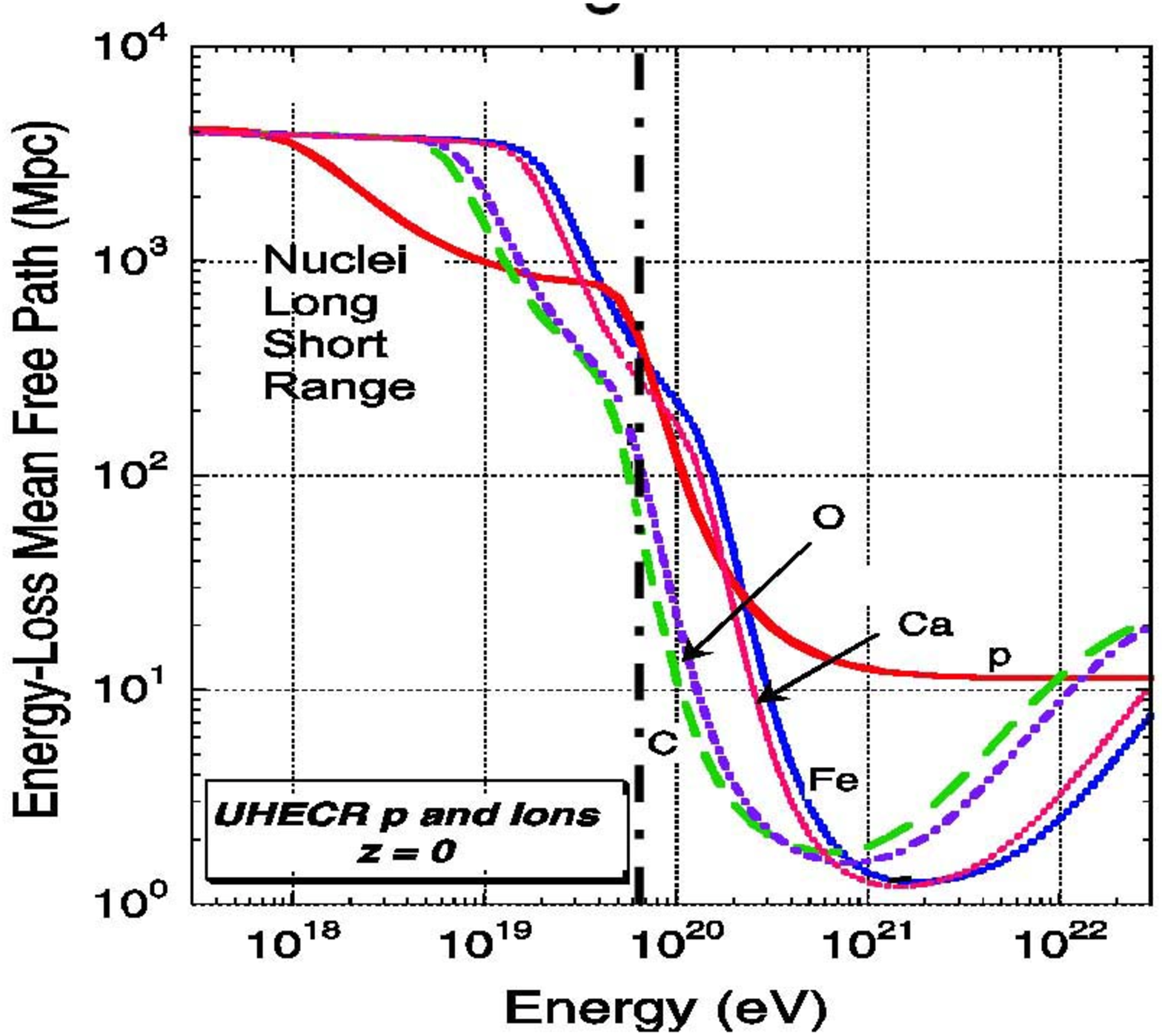}
\caption{Energy Losses and Nuclei Range:
while light nuclei are bounded in nearer Local Universe, the heavier
ones (iron nuclei) and protons are not so much suppressed at energy
$E=5.6\cdot10^{19}$ eV edge. The $p$ and $Fe$ must arrive from a wide, almost homogeneous, Universe ($R \gtrsim 500 Mpc$). Therefore, light nuclei may explain an earlier GZK cut off and the observed nearby inhomogeneity. Figure adapted from \cite{Dermer07}.}\label{fig:Heavy-p-O-Dermer}
\end{center}
\end{figure}

Last AUGER report \cite{Auger-Nov07} confirmed in more recent
\cite{Auger-Dec07} paper surprised us by its conflicting fragments
in the growing UHECR puzzle. The expected GZK cut-off took place, as
in HIRES \cite{Hires-Jun07} spectra, at too earlier energy edges ($6
\cdot 10^{19}$ eV) than those one expected for protons or iron ($1-2
\cdot 10^{20}$ eV), in order to be confined within a local Universe ($100 Mpc$)
versus the much larger proton  ($500 Mpc$). Why protons at all, if previous composition at $40$ EeV ($12$ extreme events) was leading to heavy nuclei ($\ln A= 2.6 \pm 0.6 $)? Indeed, we note here, there are other UHECR candidates whose \emph{GZK} cut off occur at earlier energies (fitting an earlier cut off), but  are nuclei: they are the light ones, which fast photo-dissociation binds
them in low energy (of few tens EeV) and nearby volume (ten or few tens Mpc),
keeping partial directionality. Their allowed volumes are therefore quite local, as shown in figure \ref{fig:Heavy-p-O-Dermer}, compatible with Cen-A spread group. This AGN source, because of the distance, is also the brightest. Much farer sources are diluted by distance and suppressed at largest edges by light nuclei cut-off. The absence itself of the rich nearby Virgo AGN is indeed still puzzling \cite{Gorbunov07}: its presence might be already hidden in earlier AGASA and Haverah Park \cite{Stanev95}; let us note that a marginal signal ($3$ events) from
 near Fornax cluster arose already. A Virgo comparable one is awaited.  In our view the very dominant presence of a much nearer Cen-A AGN source coincident with a few doublets (or even a multiple dozen
clustering at wider solid angles) suggest a key role for nearby
sources over distant ones; but Virgo and M87, I believe, should rise too, probably spread.   Why (following AUGER) the apparent farer sources from Cen clusters or even the Shapley Concentration at $z\simeq 0.02-0.04$ should shine along Super-galactic (SG) plane?\emph{"It is worth noting, as is clearly visible in Figure 2, the striking alignment of several events close to the super-galactic plane.\cite{Auger-Nov07}"} Why Cen cluster signals arise while Norma ones are not much present?  We believe that this \emph{apparent} correlation between the Super-galactic Plane (around nearest Cen-A) and the UHECR $56$ EeV events, led the AUGER collaboration to associate these to the AGN within $80-120$ Mpc. The reason for the probable blunder lays just in the rare Cen-A source position, whose dominant  emission (because of the distance) could shine more and  get spread (because of galactic magnetic fields and because nuclei composition) into a \emph{nuclei spectroscopy} along the same SG plane: the UHECR are bent by random galactic magnetic lines orthogonal to the Milky-Way disk. Moreover the lightest nuclei ($He,Be,B$) much smaller  propagation volumes makes easier to explain the presence of Cen-A (4 Mpc) and the present absence of Virgo signals (20 Mpc). This subtle coincidence between nearest Cen-A, galactic field bending and overlapping on SG plane had (probably) misled to a first  solution \cite{Auger-Nov07, Auger-Dec07}.

\begin{figure}[t]
\begin{center}
\includegraphics[width=9cm, height=6cm]{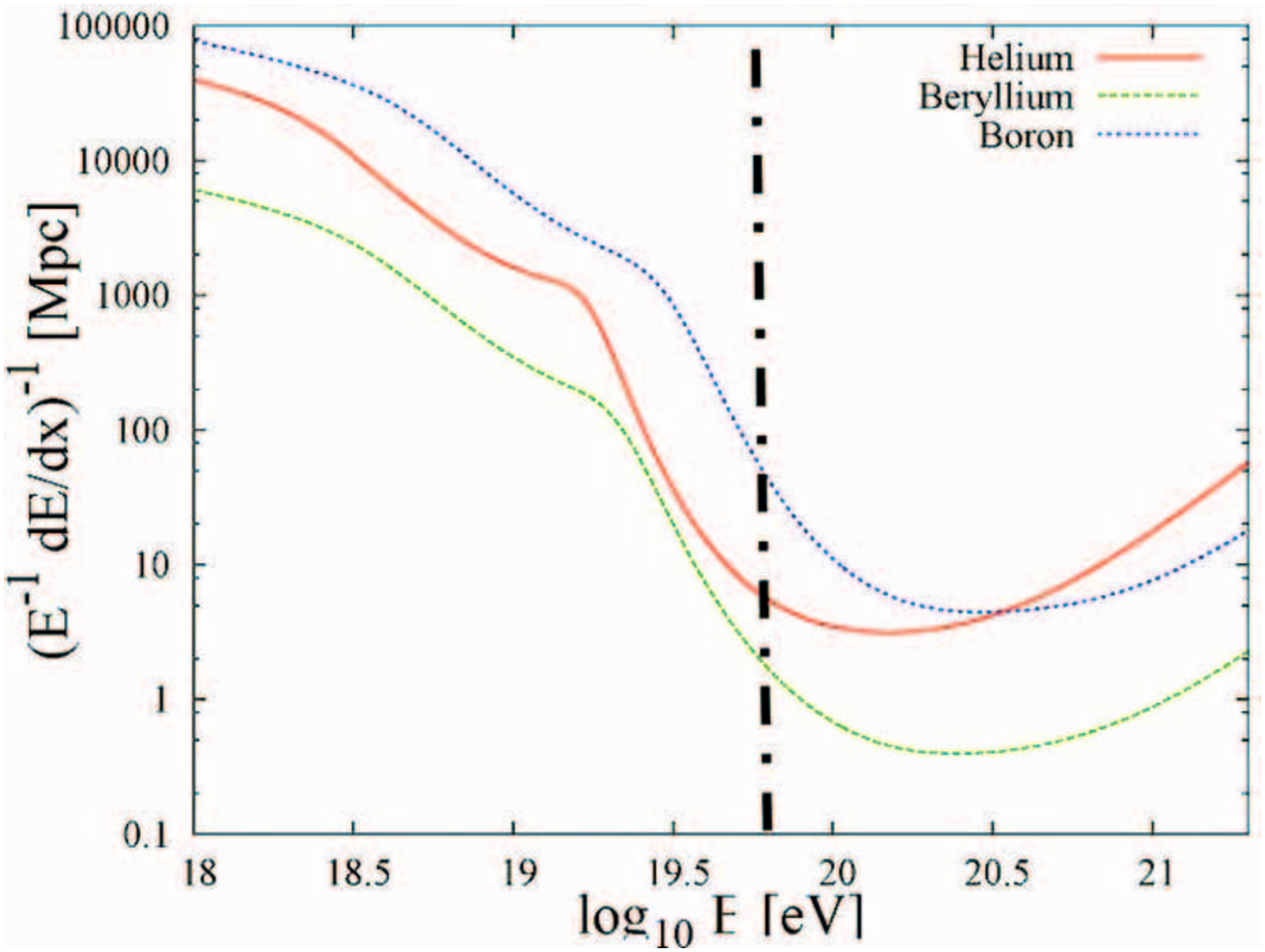}
\caption{Energy Losses and lightest nuclei
ranges (adapted from \cite{Hooper}): The Beryllium, Helium and Boron
are among the lightest nuclei that suffer a more drastic cut off
(see figure above) bounding them at nearest cosmic volumes: This may
better explain their allowed arrival from Cen-A ($4 Mpc$), their
collimation few degrees from the source and the apparent absence
from more distant Virgo ($16 Mpc$). Three events from Fornax may still
reconcile with heavier (C,O) nuclei. A single bent event from Virgo may  be present in large galactic latitude.}\label{fig:LightNuclei}
\end{center}
\end{figure}

Naturally there may be also other UHECR able to travel at
largest ranges. These are heavier ones, or the proton itself which
losses are ruled by photo-pion productions: nevertheless the nearly
absence of Virgo and the absence of more diffused noise suggest a
few or none presence of $Fe$ and $p$ candidate. At present stage and at $60$
EeV edges, we believe the sky UHECR is ruled by Cen-A light nuclei
and a few spread sources around.

The galactic magnetic field are organized in spiral way laying in
the galactic disk. The lines are frozen inside the galactic plane.
The consequent Lorentz forces are orthogonal, with opposite sign, to
it. Consequently, the lines are placed left-right in the plane; the
bending of the UHECR charges takes place up and down in the vertical
way, (as shown in figure \ref{fig:Random}) in a random way, (filling
the super-galactic arm). The average deflecting angle is
approximately:
\begin{equation}
\delta_{rm} \gtrsim {1.33^\circ}\cdot Z\frac{6\cdot10^{19}
eV}{E_{CR}}\frac{B}{\mu G}\sqrt{\frac{L}{10 kpc}}
\sqrt{\frac{l_c}{kpc}}
\end{equation}

\begin{figure}[t]
\begin{center}
\includegraphics[width=9cm, height=4cm]{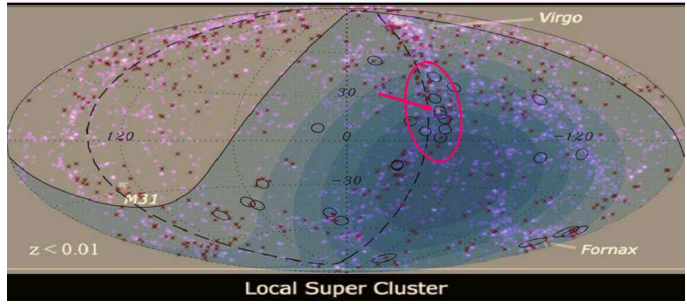}
\caption{UHECR event in nearest (redshift
$z<0.01$) Local Universe Map shows Virgo cluster absence, though
Fornax cluster is mildly observable by few events. This
contradicting argument has been well underlined recently
\cite{Gorbunov07}. The Cen-A position is marked by the arrow, while
the ten events around are within the red oval. They could be the
spread signals by UHECR bent by random galactic fields and by
different nuclei composition. The red oval clustering overlap with SuperGalactic
Plane led to a possible miss-interpretation with AGN at $80$ Mpc
in far supergalactic volumes.}\label{fig:Nearest}
\end{center}
\end{figure}

This value is respectively for He, Be, B, C, O: $
\delta_{^2He}\gtrsim{2.7^\circ}$; $\delta_{^4Be}\gtrsim{5.3^\circ}$;
$\delta_{^5B}\gtrsim{6.7^\circ}$; $\delta_{^6C}\gtrsim{8^\circ}$;
$\delta_{^8O}\gtrsim{10.7^\circ}$. The values are only approximated
and might be enhanced by a factor  probably above $2-3$. Indeed the galactic magnetic field on the plane is quite larger (three-four times at least) than the
halo one and the distances from Cen-A are crossing twice the galactic size; therefore
\begin{equation}
\delta_{rm} \gtrsim {3.76^\circ}\cdot Z\frac{6\cdot10^{19}
eV}{E_{CR}}\frac{B}{4\cdot \mu G}\sqrt{\frac{L}{20 kpc}}
\sqrt{\frac{l_c}{kpc}}
\end{equation}
This value is respectively for He, Be, B, C, O: $
\delta_{^2He}\gtrsim{7.5^\circ}$; $\delta_{^4Be}\gtrsim{11.2^\circ}$;
$\delta_{^5B}\gtrsim{18.9^\circ}$; $\delta_{^6C}\gtrsim{22.6^\circ}$;
$\delta_{^8O}\gtrsim{30.2^\circ}$. \emph{In this view it may well be possible that most of the UHECR are not all the light nuclei, but just the lightest ones (He,Be), whose propagation distance is just smaller than the Virgo distance.}
This solution may explain at best the puzzling absence of a Virgo signal.
 These bending angles are indeed well compatible with the observed angular
spread (see oval in figure \ref{fig:Nearest}) of the UHECR around Cen-A. The
most heavy iron nuclei is widely spread: $\delta_{26^Fe}\gtrsim{33.80^\circ} $ , or $\delta_{26^Fe}\gtrsim{95^\circ} $ loosing most of the arrival-source link. The absence of diffused events disfavor such a composition, contrary to \cite{Auger-Nov07} iron-proton hybrid composition assumption. At  lower energies the bending of heavy nuclei will pollute and spread homogeneously the UHECR (ten or a few tens EeVs) map. Therefore just a few light nuclei may spread the main clustering group (almost a dozen) around Cen-A.  We do not consider here the less relevant extragalactic magnetic bending because our main proposal leads to a
very Local source volume, mostly ruled out by nearby galactic last
deflections.

\begin{figure}[t]
\begin{center}
\includegraphics[width=10cm, height=4cm]{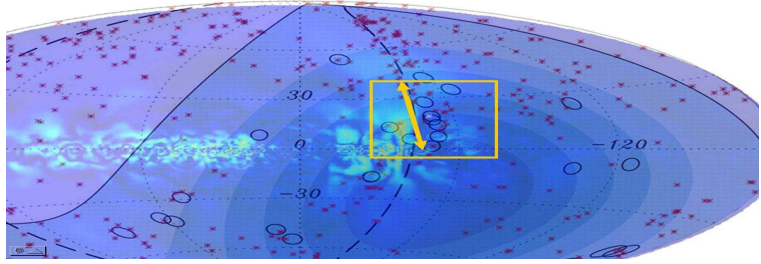}
\caption{The 2.4 GHz radio polarization,
observed by Wmap  over the UHECR AUGER events: the polarization is due to
interstellar charges and galactic magnetic lines. These signatures
imply spiral magnetic line morphology (see figure \ref{fig:Random}).
These lines also would lead to UHECR deflections  orthogonal to the
galactic disk for charged nuclei emitted by Cen-A.}\label{fig:Map-Pola2}
\end{center}
\end{figure}

\begin{figure}[t]
\begin{center}
\includegraphics[width=9cm, height=5cm]{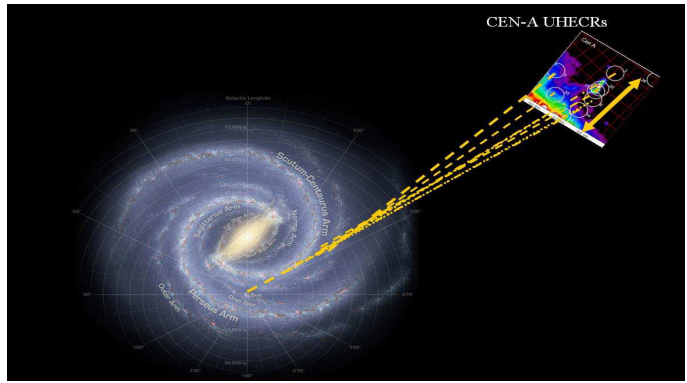}
\caption{The light and heavy nuclei may spread
up and down by random galactic magnetic field deflections. Because
of the random nature the horizontal magnetic fields act with Lorentz
forces in vertical axis bending up and down UHECR by several
degrees, depending on composition and energy.}\label{fig:Random}
\end{center}
\end{figure}

\section{Testing the present solution}
Our solution foresee that the UHECR near our Earth are ruled by
nearby source Cen-A: its shining (as observed) comes from nearest distance and by
light nuclei courier, whose small GZK cut-off  makes them bounded in
their local source nearby (in this way explaining the early energy
GZK steepness). The vertical spread (in galactic coordinate) took
place, as shown above, because of the disk horizontal spiral
magnetic lines (figure \ref{fig:Random}). Incidentally this axis
overlap part of the Super-galactic plane, leading to some confusion.
The UHECR at $60$ EeV are dominated neither by protons nor by iron
whose larger ranges would naturally offer a clearer trace of nearby
(Virgo) or a little more far Universe. AUGER present UHECR events,
we believe, are blazed by light nuclei, or even lightest ones, probably secondaries of heavier ones. The fragmentation (or even the inner AGN Jet nucleo-synthesis)  occurs possibly by photon-dissociation taking place near the AGN jet source
via self light interactions. At $40$ EeV the heavy nuclei and
protons may still be present and pollute the Universe isotropically at the
tens EeV spectra, being traveling from wider Universe volumes. At
highest energy, above $60$ EeV, we believe mostly or just  light or even lightest
nuclei may arrive. The nearly on-axis event from Cen-A ($2-5^\circ$) are
(probably) the imprint of lightest ones (He), while the more
spread events at larger angles ($8-10^\circ$) maybe the secondaries
(Be, B), whose propagation range is nevertheless much bounded than
proton or iron one, explaning at best the Virgo missing. Consequently we foresee  the crowding of future events in a  cluster \emph{vertically around} the Cen-A source  being it the real barycenter of the UHECR spread group, and no longer the far away SG plane. This model explains the lack of Virgo (\emph{whose signal might nevertheless rise soon}): indeed a single event not far from Virgo could be a very deflected $ Be, B$ one or rare $C,O$ nuclei. The model agrees with a modest signal of Fornax cluster. Rarest single  iron UHECR or protons, but even better  $C,O$ nuclei may be responsible of a few remaining spread events. The very rare overlapping doublet below the galactic disk may be related to a nearby source. The consequent signature of our proposal, compatible  to one offered just recently \cite{Wibig07}, is that events at the extreme random angles ($\delta_{rm}\simeq8-10^\circ$) far from Cen-A  must share a heavier composition  than the more collimated ones ($\delta_{rm}\simeq2-5^\circ$), being lighter (for comparable energy). This must be manifest by their airshower elongation value: $\ln A_{He}=1.38$,
$\ln A_{Be}=2.2$, $\ln A_{B}=2.38$; as we mentioned we also consider eventually $\ln A_{C}=2.485$ and $\ln A_{O}=2.77$. These values are well compatible with the AUGER claim at $40$ EeV of a dominant heavy nuclei composition  ($\ln A=2.6\pm0.6$). These values differ drastically from $\ln A_{p}=0$ for proton or $\ln A_{Fe}=4$ implied by \cite{Auger-Dec07} claim and they might be soon
tested in the UHECR length trace (slant depth) observed at best in
FD (Fluorescence Detectors). Also most deflected events from Cen-A
(at same energy range) should exhibit richer muon composition in SD
(Surface Detectors) than the less deflected ones in axis to Cen-A.

\section{Conclusions}
The AUGER discover of UHECR anisotropy has been a great achievements
which need a longer  time record. Its interpretation was very
probably hurry up and confused because an accidental coincidence  between the
Super Galactic contour and the  galactic Lorentz force bending of
light nuclei. If the AUGER interpretation ($R_{GZK}\simeq80 $ Mpc
range) is true than the expected UHE neutrino secondaries fluency (from
observed UHECR energy fluency $\phi_{UHECR_{GZK}}\simeq1 eV cm^{-2}
s^{-1}sr^{-1}$) extended to the whole Universe size
($R_{Hubble}\simeq 4 Gpc $ range) would already reach detectable
values: indeed the secondary $\phi_{\nu_{\tau}}$ energy fluency
approximate to $$\phi_{\nu_{\tau}}+\phi_{\bar{\nu_{\tau}}}\simeq\frac{1}{6}
\cdot\phi_{UHECR_{GZK}}\cdot\sqrt{(\frac{R_{Hubble}}{R_{GZK}})^3}$$
$$\phi_{\nu_{\tau}}+\phi_{\bar{\nu_{\tau}}}\simeq  60 eV cm^{-2}
s^{-1}sr^{-1} $$
This value enhanced by redshift power factor might imply a fluency
at the edge (or above) of AUGER bound \cite{Auger-Dec07-tau, Fargion07-1}. Also some EeVs gamma showers  from SG plane might already being clustering and recorded by AUGER.  Our proposal is somehow of minor impact for Neutrino Astronomy, offering a lower GZK rate, but it clarify the role of lightest nuclei in nearby UHECR
astronomy (mostly from our main Cen-A source). The consequence in UHECR
neutrino astronomy is nevertheless relevant: Cen-A might be soon
become a main major UHE neutrino source to be observed (with some
difficulties, depending on the exact source photo-pion and photo dissociation)
 by AUGER future records via Horizontal air-showers \cite{Fargion07-1,Auger-Dec07-tau}, induced by EeV UHE $\nu_{\tau}$, via their secondary  ${\tau}$ decay in air, and by their final horizontal airshower.
As well as by Magic telescope  \cite{Fargion07-2}, looking for such skimming blazing $\nu$ airshowers at horizons edges. Because of the lower UHE neutrino secondaries energy expected from lightest nuclei dissociation, the UHE tau air-showers will be better revealed at tens-hundred PeVs range in
AMIGA smaller size array detector and-or by high elevation telescopes (HEAT) to be deployed  in future AUGER  inner (Coihueco) enhanced area \cite{Auger-Oct07-tau}.

\subsection{Acknowledgments}
The author wish to thank Pietro Oliva for reading the article.
This article is devoted to the memory of Stefano Gaj Tache,
tragically lost in front of Rome Synagogue at age $2$, on 1982.

\end{document}